\documentclass{article}

\usepackage{spconf}
\usepackage{times}
\usepackage{latexsym}

\usepackage{times}
\usepackage{latexsym}
\usepackage{CJKutf8}

\usepackage[utf8]{inputenc}
\usepackage{graphicx}

\usepackage{inconsolata}

\usepackage[T1]{fontenc}

\usepackage[utf8]{inputenc}

\usepackage{microtype}

\usepackage{math-common}
\usepackage{makecell}

\usepackage{xcolor}
\usepackage{inconsolata}
\definecolor{chromeyellow}{rgb}{1.0, 0.65, 0.0}

\newcommand{\nop}[1]{}

\usepackage[T1]{fontenc}

\usepackage[utf8]{inputenc}

\usepackage{microtype}

\usepackage{inconsolata}

\usepackage{colortbl,booktabs}
\definecolor{mygray}{gray}{0.95}

\usepackage{amssymb}
\usepackage{pifont}

\usepackage{numprint,siunitx,xcolor}
\usepackage{multirow}
\usepackage{tabu}
\usepackage{graphics}
\usepackage{float}

\renewcommand{\paragraph}[1]{\noindent\textbf{#1}}

\title{On the Open Prompt Challenge in Conditional Audio Generation}
\name{Ernie Chang$^{\spadesuit}$, Sidd Srinivasan$^{\spadesuit}$, Mahi Luthra$^{\spadesuit}$, Pin-Jie Lin$^{\clubsuit}$, Varun Nagaraja$^{\spadesuit}$, Forrest Iandola$^{\spadesuit}$,}
\myname{Zechun Liu$^{\spadesuit}$, Zhaoheng Ni$^{\spadesuit}$, Changsheng Zhao$^{\spadesuit}$, Yangyang Shi$^{\spadesuit}$ and Vikas Chandra$^{\spadesuit}$}

\address{$^\spadesuit$Meta AI\\
         $^\clubsuit$Language Science and Technology, Saarland University\\
         \texttt{\{erniecyc, siddsrinivasan, mahiluthra\}@meta.com} \\
         \texttt{pinjie@lst.uni-saarland.de}}

\begin{document}
%
 \maketitle
\begin{abstract}

Text-to-audio generation (TTA) produces audio from a text description, learning from pairs of audio samples and hand-annotated text. 
However, commercializing audio generation is challenging as user-input prompts are often under-specified when compared to text descriptions used to train TTA models. 
In this work, we treat TTA models as a ``blackbox'' and address the user prompt challenge with two key insights:
(1) User prompts are generally under-specified,  leading to a large alignment gap between user prompts and training  prompts.
(2) There is a distribution of audio descriptions for which TTA models are better at generating higher quality audio, which we refer to as \emph{``audionese''}.
To this end, we rewrite prompts with instruction-tuned models and propose utilizing text-audio alignment as feedback signals via margin ranking learning for audio improvements. 
On both objective and subjective human evaluations, we observed marked improvements in both text-audio alignment and music audio quality.


\end{abstract}

\section{Introduction}


Text-to-audio (TTA) generation has witnessed significant advancements in recent years, enabling the conversion of textual descriptions into high-fidelity audio representations~\cite{yang2023diffsound,kreuk2022audiogen}. 
TTA models have been trained using paired data consisting of hand-annotated texts and corresponding audio samples, leveraging neural approaches to learn the mapping between text and audio.

Despite these advancements, scarcity in paired text-audio data has created an inherent difficulty in synthesizing high-quality and coherent audio from text. Creating text descriptions of general audio is considerably harder than describing images~\cite{ramesh2021zero}.
MusicLM \cite{agostinelli2023musiclm} outlines two challenges in creating music prompts:
(1) Expressing the essential features of acoustic scenes 
and music 
is a complex task that cannot be easily accomplished using only a few words.
(2) The temporal dimension in audio introduces a structural complexity that renders sequence-wide captions less effective as annotations compared to image captions.

\begin{figure}
  \centering
\includegraphics[width=0.8\columnwidth]{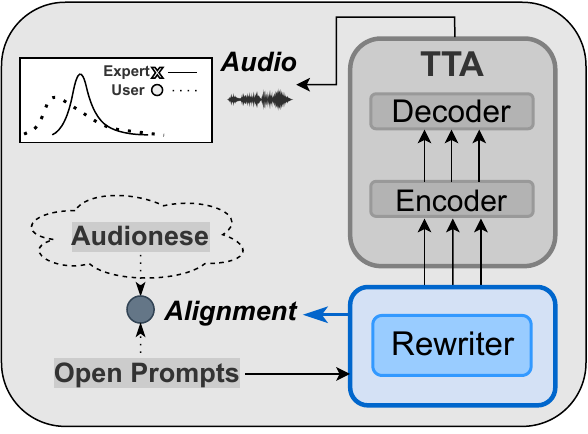}
\caption{ \small
\textbf{Alignment of open prompts:}
Our approach involves an LLM-based prompt rewriter to align user (open) prompts to ``audionese'', which is a distribution of text capable of producing higher quality audio given the blackboxed text encoder and audio decoder. 
}
\label{fig:overview}
\end{figure}



The problem is exacerbated when conditional audio models are put into actual product usage -- it results in misalignment\footnote{This train-test deviation has long been studied in past alignment research~\cite{Schick2022PEERAC,Yasunaga2020DrRepair,Welleck2022SelfCorrect,bai2022constitutional}. 
} between user prompts and annotated prompts, making it difficult for audio decoders to synthesize audio samples that accurately represent the user intents.
Such prompt alignment is worse in musical domains -- in line with previous studies \cite{agostinelli2023musiclm}, we observed that majority of users lack the expertise to construct musical prompts that are as descriptive as those in the training set (which are typically annotated by domain experts). 
Therefore, (open) user prompts are often too abstract and underspecified (e.g., \emph{``pop song''}) compared to the annotated, elaborate prompts that a TTA model is typically trained on (e.g., \emph{``Moody keyboard and drum centric pop song featuring neo-R\&B chordal information and layered barbershop harmonies''}).

In this work, we provide a preliminary study of the open prompt challenge in text-to-audio generation where we expose TTA models to out-of-distribution user text prompts. We focus our efforts on musical audio (as opposed to speech or other general acoustic stimuli) because of the aforementioned gap between general user and expert prompts for music.
We leverage instruction-tuned large language models (LLMs) to perform prompt rewriting and improve upon the LLMs with audio feedback and margin rank learning to increase their ability to output prompts capable of producing higher quality audio.
Overall, we observe improvements over audio quality, text-audio alignments, and human preference.

\label{sec:intro}

\section{Open Prompts Collection}

Text-to-audio (TTA) models generate audio from a text description $x$ such as 
\textit{``The pop rock music features a male voice singing.''}
TTA systems are designed to generate a wide range of high dimensional audio signals $y$ by modeling the learned compact latent space $Z$.
Text encoders of TTA systems condition on $x$ to sample latent code $z$, which the audio decoder uses as the prior to sample $y$. 
This process hinges on the alignment between the learned parameter $\theta$ of both text encoder and audio decoder. 
Thus, the problem of prompt rewriting targets at converting text inputs $x$ into $x'$ such that they are closer to the \emph{audionese} and can thus better leverage $\theta$ to produce high-fidelity audio samples (refer Figure~\ref{fig:overview}). 
Here we define audionese as the text distribution that produces the best audio quality metrics for a given TTA model parameterized by parameters $\theta$, which only exists theoretically as a result of the model training and has its root in input perturbation and model complexity~\cite{novaksensitivity}.

\paragraph{Out-of-distribution prompts.} For our study, we set up collections runs with $30$ non-expert users with varying knowledge about music. 
We collected $300$ user input texts by asking users to enter free-form text prompts for generating music.
This resulted in a wide range of musical prompt topics, many of which did not belong to musical domains (e.g., we found prompts relating to cuisines, sports, and politics.)



\paragraph{Expert vs. user prompts.} We compared the linguistic complexity of expert and open prompts, by examining their distributional difference along 
information density (or entropy) $log (\frac{1}{ P(w_j | t_i)})$~\cite{demberg2008data}. Here, $w_j$ is $j^{th}$ token and $t_i$ is the set of tokens in prompts. 

\begin{figure}[h]
  \centering
\includegraphics[width=1\columnwidth]{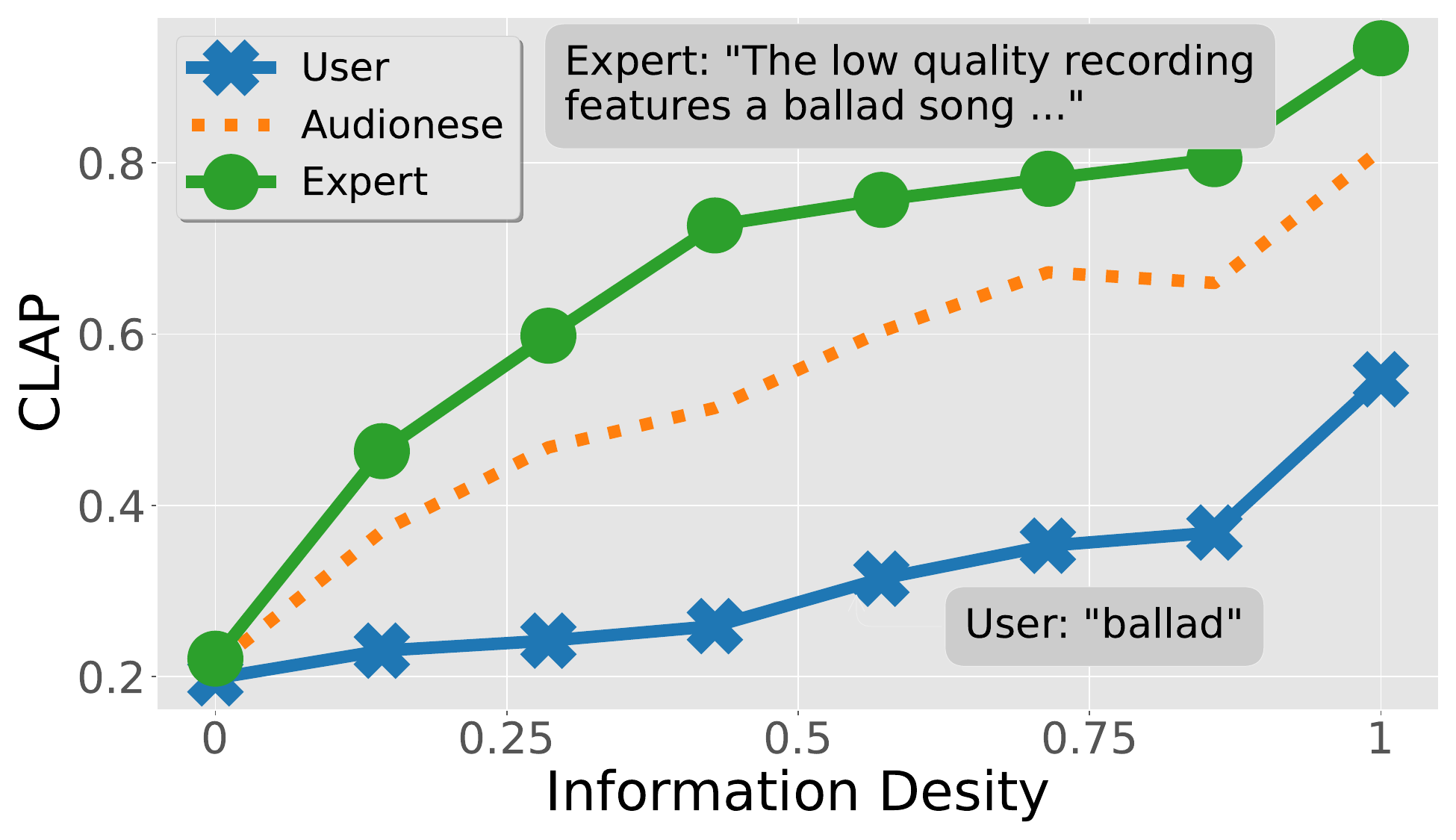}
\caption{ \small \textbf{Comparison of user and expert prompts:}
Plot of normalized CLAP and information density scores where audio descriptive prompts range from expert-annotated musical prompts to user-input prompts, which subsequently result in various text-audio alignment (CLAP). 
The disparity is due to difference in ease of annotation where expert-annotation are costly to obtain as opposed to arbitrary open prompts.
The dotted line represents processed versions of the user prompts, rewritten by a finetuned LLM with the intent to push them closer to the \emph{audionese}. 
}
\label{fig:dist}
\end{figure}


\begin{table*}[h]
    \centering
    \resizebox{\textwidth}{!}{
    \begin{tabu}{p{0.15\linewidth} p{0.4\linewidth}p{0.45\linewidth}}
\toprule
\textbf{User Input} & \textbf{ Instruction} & \textbf{Enhanced Prompt} \\
\midrule
\emph{Cupcake}

& \emph{Write a detailed description describing the music with instruments, tone, pitch, rhythm, atmosphere, style, and all the audio details.}

& \emph{The music for the cupcake video is upbeat and cheerful, featuring a mix of bouncy electronic elements and live instruments such as guitar and drums. The tone is light and playful, with a catchy melody and a steady beat that evokes feelings of happiness and joy.} \\

\midrule
\emph{Write a detailed description describing the music with instruments, tone, pitch, rhythm, atmosphere, style, and all the audio details.}

& \emph{Rewrite the following prompts to be diverse and different}

& \emph{1. Describe the instruments used in the music, including their tone, pitch, and rhythm. 2. Provide details about the overall sound and atmosphere of the music, including the style and genre. 3. Include any special effects or layers that enhance the music, such as layers of guitars or synths.} \\
\bottomrule
\end{tabu}
}
\caption{\small The provided example represents a pair of instruction-input and output generated by instruction-tuned LLMs. Here, the LM must infer a latent document-level concept to generate
coherent next tokens during pretraining. At test time, in-context learning occurs when the LM also infers
a shared latent concept between examples in a prompt.}\label{tab:example}
\end{table*}

Further, we use the CLAP metric ~\cite{elizalde2023clap} as a proxy for alignment between text and audio pairs. 
CLAP consists of language and audio encoders trained with a contrastive learning objective to project audio and text descriptions into the same latent space and can thus serve as a useful estimate of their alignment. To verify this, we also conducted a quick study to correlate the CLAP metric to human evaluations of text-audio alignment, finding significant correlations at (\textit{r=0.35, p<.05}). 

We then conducted analyses comparing CLAP scores to informational density of the text prompt, summarized in Figure~\ref{fig:dist}:
(A) Relationship between information density and CLAP is observed to be monotonically increasing, which suggests a correlation between how much information is packed within each prompt and how well it aligns with the generated audio.
(B) There is a large disparity between expert and open prompts for both CLAP and information density. Further analysis shows that open prompts not only use a smaller vocabulary but the average prompt length is also lower by $5$ tokens.





\label{sec:format}

\section{The Approach}
\label{sec:approach}

We employ instruction-tuned large language models, i.e. FLAN-T5 variants~\cite{chung2022scaling} and LaMini-LM~\cite{wu2023lamini} as base models $f(\cdot)$ (\textbf{\emph{Instruct}}), and improve upon them with audio feedback in the form of text-audio ranking. 
Each prompt $x$ is passed into base model $f(x)$ to obtain a converted prompt $x'$.  
Since the base models are instruction-tuned, they possess generic instruction capabilities that enabled them to rewrite user prompts with some level of audio details. We employ this approach as the zero-shot approach (\textbf{\emph{Instruct+0-shot}}) of rewriting prompts.
However, the ability for these models to converse in \emph{audionese} without any finetuning is still largely limited.
In what follows we describe how supervised update is performed to achieve this.

\subsection{Open Prompt Rewriting with Feedback Alignment}
\paragraph{Step 1: Audio feedback.} As earlier, we use the CLAP score as a pointwise metric to evaluate text-audio alignment for each pair by computing a score in the range $[0,1]$. We first generate a diverse set of enhanced prompts $x'$ by employing the zero-shot LLM approach (described above) on user prompts. This thus gives us pairs of the original user prompt $x$ and the rewritten prompt $x'$ and CLAP scores of each which can then be used for ranking the items in each pair.

\paragraph{Step 2: Rank-based alignment.} For \textbf{\emph{Instruct+Feedback}}, 
we introduce the margin ranking loss, which evaluates two types of prompts, denoted as $x$ and $x'$. 
The variable $y$ represents the ranking expectation, where 
a value of 1 for $y$ indicates the desired ranking order where $x$ is ranked higher than $x'$. Conversely, a value of -1 indicates the desired ranking order $x'$ higher than $x$.
The loss is computed using the formula: $\max(0, -y \cdot (x - x') + m)$, where $m$ represents the margin, a constant determining the minimum difference required between the ranked values to avoid incurring a penalty. 
If the difference between $x$ and $x'$ falls below the margin, indicating a satisfactory ranking, then the loss is assumed to be $0$ to signify its desired property.
See sample generation at Table~\ref{tab:example}.


\begin{table}[h]
    \centering
    \resizebox{\columnwidth}{!}{
    \begin{tabu}{l|l}

\tabucline [2pt]{1-2}
\noalign{\vskip 1mm}
\huge\textbf{User Input} & \huge\textbf{Enhanced Prompt (Instruct+Feedback)} \\
\midrule

\noalign{\vskip 2mm}
\huge\texttt{bumpy jazz}	& 
\huge\texttt{The music is bumpy and jazzy} \\ \hline

\noalign{\vskip 2mm}  
\huge\texttt{sad harp solo}	& 
\huge\texttt{The harp solo is sad and haunting} \\ \hline

\noalign{\vskip 2mm}
\huge\texttt{cozy warm hip hop beat}	&
\huge\texttt{a hip hop beat with a warm atmosphere} \\\hline

\noalign{\vskip 2mm}
\huge\texttt{fashion upbeat}	&
\huge\texttt{The music is upbeat and energetic} \\\hline

\noalign{\vskip 2mm}
\huge\texttt{gnarly beats that are fast and crazy}	&
\huge\texttt{a sonic journey} \\ 
\noalign{\vskip 1mm}
\tabucline [2pt]{1-2}
\end{tabu}}
\caption{\small The provided example represents a pair of instruction-input and output that is utilized to train instruction-tuned LLMs.}\label{tab:example}
\end{table}

\label{sec:pagestyle}
\begin{table*}[t]
\centering
\resizebox{\textwidth}{!}{
\begin{tabular}{l|ccccc}
\toprule
             & \multicolumn{1}{l|}{LaMini-LM (125M)} & \multicolumn{1}{l|}{LaMini-LM (1.5B)} & \multicolumn{1}{l|}{FLAN-T5-small (80M)} & \multicolumn{1}{l|}{FLAN-T5-base (250M)} & \multicolumn{1}{l}{FLAN-T5-large (780M)}  \\ \hline
Open Prompts & \multicolumn{5}{c}{\cellcolor{mygray}0.0556 (100.00)}       \\ \hline
Instruct & \multicolumn{1}{c|}{0.0123 (1.58)}       & \multicolumn{1}{c|}{0.0232 (3.27)} & \multicolumn{1}{c|}{0.0744 (56.61)}     & \multicolumn{1}{c|}{0.0680 (61.20)}              & \multicolumn{1}{c}{0.0701 (30.50)}                                              \\ \hline
 Instruct+0-shot & \multicolumn{1}{c|}{0.0175 (1.88)}       & \multicolumn{1}{c|}{0.0299 (4.31)}  & \multicolumn{1}{c|}{0.0747 (56.68)}    & \multicolumn{1}{c|}{0.0693 (62.74)}              & \multicolumn{1}{c}{0.0763 (31.71)}                                             \\ \hline
 Instruct+Feedback    & \multicolumn{1}{c|}{0.126 (1.32)}       & \multicolumn{1}{c|}{0.115 (2.59)} & \multicolumn{1}{c|}{0.0739 (55.01)}     & \multicolumn{1}{c|}{0.0769 (27.96)}              & \multicolumn{1}{c}{0.0809 (29.46)}                                          \\ 

\bottomrule
\end{tabular}}
\caption{ \small Benchmarks showing the CLAP  scores for various approaches including the collected \emph{open prompts}, \emph{Instruct} with \emph{0-shot} and \emph{Instruct+Feedback}. 
We indicate SacreBLEU (\%) with original prompt as reference to show the extent of textual deviation.
We also compute the CLAP scores with the audio samples generated from test user prompts to show the theoretical lower bound.
}
\label{tab:main}
\end{table*}


\section{Experiments and Results}


\paragraph{Configs.} In this work, we employ AudioLDM~\cite{liu2023audioldm} to generate realistic speech and piano music audio samples. 
AudioLDM uses the CLAP model~\cite{elizalde2023clap} as the text encoder to obtain the text embeddings and applies a diffusion model to predict the quantized mel spectrogram features of the target audio. 
We avoid using LLMs larger than 1.5B due to resource contraints as per realtime inference.
Thus, we use LaMini-LM-[125M, 1.5B]~\cite{wu2023lamini}, which are decoder-only language models; and FLAN-T5 are encoder-decoder based models~\cite{chung2022scaling} with sizes 250M, 780M, and up to 3B due to empirical memory and latency constraints.
Our codebase is released and built with the Huggingface library \cite{wolf2020transformers}.
Audio samples are evaluated with CLAP~\cite{elizalde2023clap} for automatic text-audio alignment and subjective/objective human evaluation~\cite{liu2023audioldm,agostinelli2023musiclm} for audio quality assessment based on human preference.
We sampled $250$ expert-annotated prompts from MusicCaps~\cite{agostinelli2023musiclm} to test for performance degradation and use the collected $300$ open prompts with $250$ as test samples, and the remaining $50$ for training.
\label{sec:typestyle}

\paragraph{Main results.} We first present the main results on Table~\ref{tab:main} where we compare \emph{Original}, \emph{0-shot-refinement}, \emph{Instruct}, and \emph{Instruct+Feedback}. 
To demonstrate the efficacy of our proposed approach, \emph{Instruct+Feedback}, we compare it against two alternative methods: \emph{Instruct+0-shot} and the base instruction-tuned LLM (\emph{Instruct}), where we observe marked and consistent CLAP-based improvement over open prompts.
We also observe that the proposed technique results in a plot resembling the hypothesized audionese, and we attribute this finding to the ability of LLMs to embellish open prompts.

\paragraph{Ablation studies.}  By leveraging the CLAP-based feedback, \emph{Instruct+Feedback} improves unseen prompts as well, suggesting its ability to project the prompt into a text distribution more attuned to the TTA models. 
In contrast, \emph{Instruct+0-shot} relies solely on pre-trained models and does not benefit from user feedback when exposed to unseen prompts. 
Overall, our experimental results consistently show that \emph{Instruct+Feedback} achieves higher CLAP scores as compared to \emph{Instruct+0-shot} and the original prompts.

\begin{table}[h]
\centering
\resizebox{\columnwidth}{!}{%
\begin{tabu}{lcc|cc}
\tabucline [1pt]{1-5}
\multirow{2}{*}{\textbf{Model}} & \multicolumn{2}{c}{\textbf{MusicCaps (Close)}} & \multicolumn{2}{c}{\textbf{User (Open)}} \\
 & \texttt{SBJ(\%)} & \texttt{OBJ} & \texttt{SBJ(\%)} & \texttt{OBJ} \\ \hline
Original              & \cellcolor{mygray}24.4  & \cellcolor{mygray}3.47 & \cellcolor{mygray}10.4  & \cellcolor{mygray}1.53 \\
Instruct               & 23.6  & 3.58 & 28.4  & 3.63 \\
Instruct+0-shot             & \textbf{26.8}  & 3.39 & 29.6  & 3.65 \\
Instruct+Feedback   			  & 25.2  & \textbf{3.47} & \textbf{31.6}  & \textbf{3.71} \\
\tabucline [1pt]{1-5}
\end{tabu}%
}
\caption{\small 
Human evaluation of generated audio samples (with FLAN-T5-large) for \emph{close} and \emph{open} prompts.
Five annotators were asked to evaluate the \emph{Subjective} (side-by-side audio preference) and \emph{Objective} (relevance to prompts on scale 1-5).
}
\label{tab: human_evaluation}
\end{table}

\begin{figure}[h]
  \centering
\includegraphics[width=1\columnwidth]{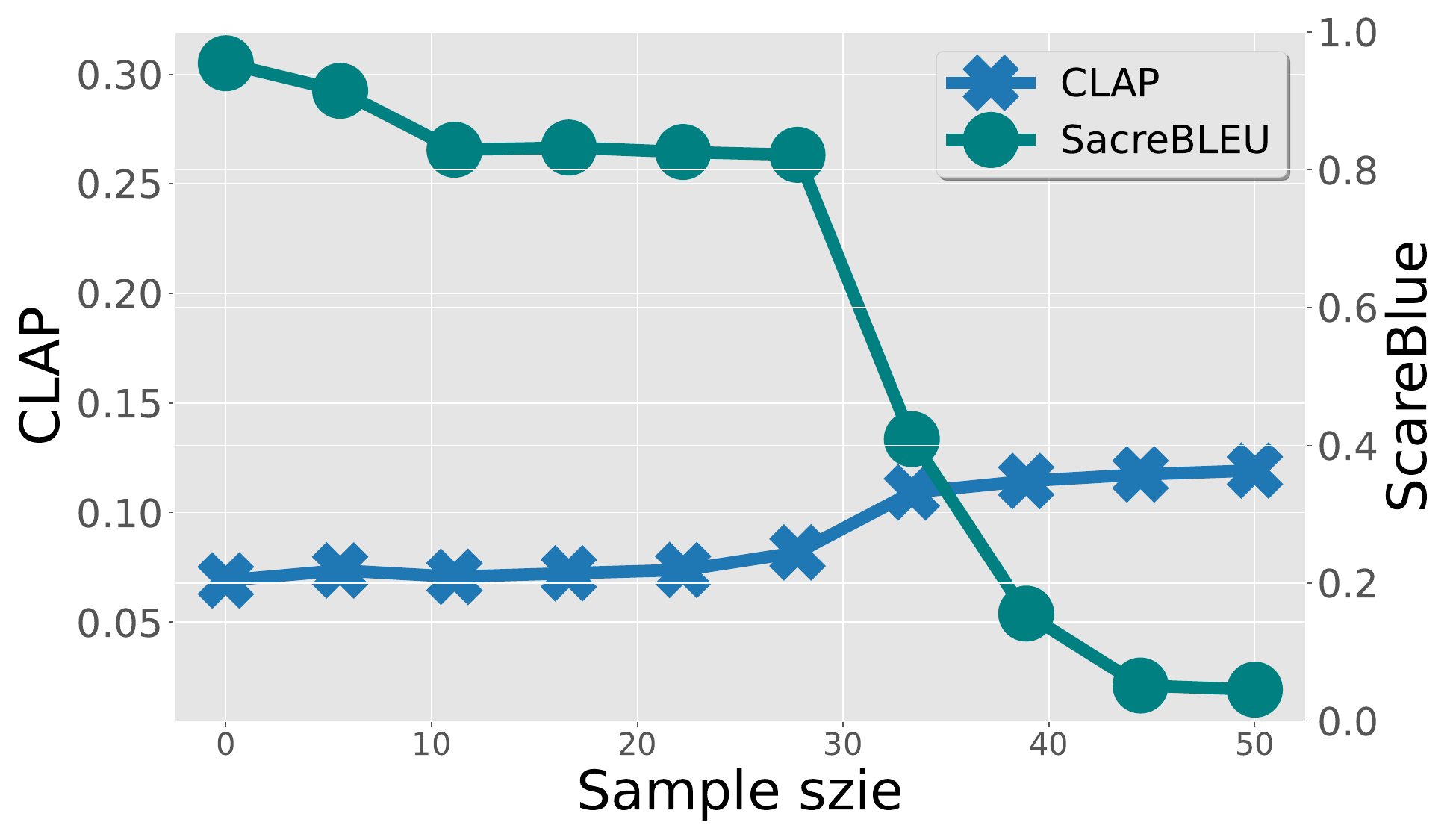}
\caption{ \small \textbf{Learning curve:}
The figure illustrates the learning curve at small sample size of $50$, with an iterval size of $5$. We report the CLAP and ScareBLEU scores.}
\label{fig:learning_curve}
\end{figure}

\paragraph{Effectiveness in low resource scenarios.} Further, while the training set consists of $50$ prompts, CLAP-based improvement was observed with as little as $5$ prompts, and gradually plateaued at $10$ samples (see Figure~\ref{fig:learning_curve}) -- this opens up the possibility of online learning of prompt rewriter models, where shifts in the distribution of user open prompts can be readily acquired.
Thus, we think that the approach is suitable for online learning setups as well, which we save for future works.

In practice, we found the margin rank learning process to be rather brittle. 
We set the learning rate to be 3e-4 and with training samples up to $50$ samples, and observed that the attained CLAP scores to go significantly higher as more samples are added, but at the huge cost of the text similarity with the original prompts. 
To avoid drastic deviation from the original user intent (and hence the objective human evaluation), we pick lower training sample sizes between $5$-$10$, depending on the SacreBLEU threshold, and stop training when SacreBLEU goes below $20$ points. 
We summarize the relationship between SacreBLEU and CLAP below in Figure~\ref{fig:learning_curve}. 

\paragraph{Choice of model architectures.}
Moreover, we observe more visible improvements with encoder-decoder based architecture as shown in the FLAN-T5 series, while decoder-only LaMini-LM seems to display better CLAP scores, but results in extremely low SacreBLEU scores, which translates to text-level degradation upon further examination.
We attribute this to the encoder's ability to more robustly encode full sequence all-at-once, thereby either mitigating the noise or accounting for the full context before decoding.



\paragraph{Human preference study.} Overall, we found that \emph{Instruct+Feedback} generates higher-quality and more contextually appropriate prompts, but the base model \emph{Instruct} offers the greater quality improvement.
The alignment technique we use is limited by CLAP's ability to effectively identify decent audio qualities, as shown the MusicCaps test set -- leading to our hypothesis that CLAP provides the greatest improvements for open, abstract prompts, rather than elaborate prompts.

\paragraph{Correlation of CLAP with human preference.} Interestingly, the gap between the preferential ratio between Open \emph{Original Prompts} and \emph{Instruct} is greater than that as measured by CLAP (Table~\ref{tab: human_evaluation}).
In terms of objective evaluation, \emph{Instruct+Feedback} is the clear winner and yields a slightly larger gap than the case of subjective evaluation.
However, we measure the strength of CLAP correlation with human preference, and found that the correlation coefficient to be a mere $0.242$ --
which is a limiting factor of our experiments, so we hope to explore with better metrics to account for audio samples' temporal information with human preference incorporated. 

\label{sec:majhead}


\section{Conclusions}
This work addresses the open prompt challenge for commercialized TTA generation by proposing the concept of "audionese" and enhancing user prompts. 
Our observation highlights vague user prompts causing alignment issues with training data. Certain audio descriptions yield improved TTA model results, named "\emph{audionese}", exposing intricacies and emphasizing the need for better alignment and \emph{audionese} comprehension.
We analyze the distribution of audio text prompts and propose prompt enhancement techniques using instruction-tuned large language models. 
Through extensive experiments, the proposed approach demonstrates significant improvements in audio metrics compared to the original user prompts, as validated by objective and subjective human evaluations.

\bibliographystyle{IEEEbib}
\bibliography{acl2021,anthology,coreset,custom,datasets,music,prompts}

\begin{thebibliography}{10}

\bibitem{yang2023diffsound}
Dongchao Yang, Jianwei Yu, Helin Wang, Wen Wang, Chao Weng, Yuexian Zou, and
  Dong Yu,
\newblock ``Diffsound: Discrete diffusion model for text-to-sound generation,''
\newblock {\em IEEE/ACM Transactions on Audio, Speech, and Language
  Processing}, 2023.

\bibitem{kreuk2022audiogen}
Felix Kreuk, Gabriel Synnaeve, Adam Polyak, Uriel Singer, Alexandre
  D{\'e}fossez, Jade Copet, Devi Parikh, Yaniv Taigman, and Yossi Adi,
\newblock ``Audiogen: Textually guided audio generation,''
\newblock {\em arXiv preprint arXiv:2209.15352}, 2022.

\bibitem{ramesh2021zero}
Aditya Ramesh, Mikhail Pavlov, Gabriel Goh, Scott Gray, Chelsea Voss, Alec
  Radford, Mark Chen, and Ilya Sutskever,
\newblock ``Zero-shot text-to-image generation,''
\newblock in {\em International Conference on Machine Learning}. PMLR, 2021,
  pp. 8821--8831.

\bibitem{agostinelli2023musiclm}
Andrea Agostinelli, Timo~I Denk, Zal{\'a}n Borsos, Jesse Engel, Mauro Verzetti,
  Antoine Caillon, Qingqing Huang, Aren Jansen, Adam Roberts, Marco
  Tagliasacchi, et~al.,
\newblock ``Musiclm: Generating music from text,''
\newblock {\em arXiv preprint arXiv:2301.11325}, 2023.

\bibitem{Schick2022PEERAC}
Timo Schick, Jane Dwivedi-Yu, Zhengbao Jiang, Fabio Petroni, Patrick Lewis,
  Gautier Izacard, Qingfei You, Christoforos Nalmpantis, Edouard Grave, and
  Sebastian Riedel,
\newblock ``Peer: A collaborative language model,''
\newblock {\em ArXiv}, vol. abs/2208.11663, 2022.

\bibitem{Yasunaga2020DrRepair}
Michihiro Yasunaga and Percy Liang,
\newblock ``{Graph-based, self-supervised program repair from diagnostic
  feedback},''
\newblock {\em 37th Int. Conf. Mach. Learn. ICML 2020}, vol. PartF168147-14,
  pp. 10730--10739, 2020.

\bibitem{Welleck2022SelfCorrect}
Sean Welleck, Ximing Lu, Peter West, Faeze Brahman, Tianxiao Shen, Daniel
  Khashabi, and Yejin Choi,
\newblock ``Generating sequences by learning to self-correct,''
\newblock {\em arXiv preprint arXiv:2211.00053}, 2022.

\bibitem{bai2022constitutional}
Yuntao Bai, Saurav Kadavath, Sandipan Kundu, Amanda Askell, Jackson Kernion,
  Andy Jones, Anna Chen, Anna Goldie, Azalia Mirhoseini, Cameron McKinnon,
  et~al.,
\newblock ``Constitutional ai: Harmlessness from ai feedback,''
\newblock {\em arXiv preprint arXiv:2212.08073}, 2022.

\bibitem{novaksensitivity}
Roman Novak, Yasaman Bahri, Daniel~A Abolafia, Jeffrey Pennington, and Jascha
  Sohl-Dickstein,
\newblock ``Sensitivity and generalization in neural networks: an empirical
  study,''
\newblock in {\em International Conference on Learning Representations}.

\bibitem{demberg2008data}
Vera Demberg and Frank Keller,
\newblock ``Data from eye-tracking corpora as evidence for theories of
  syntactic processing complexity,''
\newblock {\em Cognition}, vol. 109, no. 2, pp. 193--210, 2008.

\bibitem{elizalde2023clap}
Benjamin Elizalde, Soham Deshmukh, Mahmoud Al~Ismail, and Huaming Wang,
\newblock ``Clap: Learning audio concepts from natural language supervision,''
\newblock in {\em ICASSP 2023-2023 IEEE International Conference on Acoustics,
  Speech and Signal Processing (ICASSP)}. IEEE, 2023, pp. 1--5.

\bibitem{chung2022scaling}
Hyung~Won Chung, Le~Hou, Shayne Longpre, Barret Zoph, Yi~Tay, William Fedus,
  Eric Li, Xuezhi Wang, Mostafa Dehghani, Siddhartha Brahma, et~al.,
\newblock ``Scaling instruction-finetuned language models,''
\newblock {\em arXiv preprint arXiv:2210.11416}, 2022.

\bibitem{wu2023lamini}
Minghao Wu, Abdul Waheed, Chiyu Zhang, Muhammad Abdul-Mageed, and Alham~Fikri
  Aji,
\newblock ``Lamini-lm: A diverse herd of distilled models from large-scale
  instructions,''
\newblock {\em arXiv preprint arXiv:2304.14402}, 2023.

\bibitem{liu2023audioldm}
Haohe Liu, Zehua Chen, Yi~Yuan, Xinhao Mei, Xubo Liu, Danilo Mandic, Wenwu
  Wang, and Mark~D Plumbley,
\newblock ``Audioldm: Text-to-audio generation with latent diffusion models,''
\newblock {\em arXiv preprint arXiv:2301.12503}, 2023.

\bibitem{wolf2020transformers}
Thomas Wolf, Lysandre Debut, Victor Sanh, Julien Chaumond, Clement Delangue,
  Anthony Moi, Pierric Cistac, Tim Rault, R{\'e}mi Louf, Morgan Funtowicz,
  et~al.,
\newblock ``Transformers: State-of-the-art natural language processing,''
\newblock in {\em Proceedings of the 2020 conference on empirical methods in
  natural language processing: system demonstrations}, 2020, pp. 38--45.

\end{thebibliography}


\end{document}